\UseRawInputEncoding
\documentclass[preprint,12pt]{elsarticle}

\usepackage{graphicx}
\usepackage{algorithm,algorithmic}
\usepackage{amssymb}
\usepackage{epstopdf}
\usepackage{listings}
\usepackage{float}
\usepackage{lineno}
\usepackage{etex}
\usepackage{graphicx,color,colordvi}
\usepackage{latexsym,amsbsy,epsfig,fancybox}
\usepackage{amstext}
\usepackage{subfigure}
\usepackage{amsfonts,textgreek}
\usepackage{mathrsfs}
\usepackage{mathtools}
\usepackage{stmaryrd,dsfont}
\usepackage{bbm}
\usepackage{url}
\usepackage{wrapfig,bm}
\usepackage{tikz}
\usepackage{pstricks}
\usepackage{caption}
\usepackage{psfrag}
\usepackage{upgreek}
\usepackage{isomath}
\usepackage{latexsym}
\usepackage{array}
\usepackage{multirow}
\usepackage{booktabs}
\usepackage{verbatim}
\usepackage{natbib}
\biboptions{sort&compress}
\usepackage{color}
\usepackage{colortbl}
\usepackage[colorlinks=true,citecolor=blue]{hyperref}
\usepackage{overpic}
\usepackage{enumerate}
\usepackage{amsmath,amssymb,xspace}
\usetikzlibrary{plotmarks}
\usetikzlibrary{fit,positioning,arrows}
\usetikzlibrary{shapes.geometric, plotmarks}
\usepackage{appendix}

\usepackage[margin=3cm]{geometry}
\usepackage{soul}

\journal{Journal of Banking and Financial Technology}
\begin{document}
\begin{frontmatter}

\title{StockBot: Using LSTMs to Predict Stock Prices}

\author[1]{Shaswat Mohanty\corref{cor1}} 
\ead{shaswatm@stanford.edu}
\author[2]{Anirudh Vijay}
\author[1]{Nandagopan Gopakumar}

\cortext[cor1]{Corresponding author}
	
\address[1]{Department of Mechanical Engineering, Stanford University, CA 94305-4040, USA}
\address[2]{Department of Electrical Engineering, Stanford University, CA 94305-4040, USA}


\begin{abstract}
The evaluation of the financial markets to predict their behaviour have been attempted using a number of approaches, to make smart and profitable investment decisions. Owing to the highly non-linear trends and inter-dependencies, it is often difficult to develop a statistical approach that elucidates the market behaviour entirely. To this end, we present a long-short term memory (LSTM) based model that leverages the sequential structure of the time-series data to provide an accurate market forecast. We then develop a decision making StockBot that buys/sells stocks at the end of the day with the goal of maximizing profits. We successfully demonstrate an accurate prediction model, as a result of which our StockBot can outpace the market and can strategize for gains that are $~15$ times higher than the most aggressive ETFs in the market.

\end{abstract}

\begin{keyword}
Stock Prediction \sep StockBot \sep Long-Short Term Memory \sep Stock Decision-making Bot
\end{keyword}

\end{frontmatter}

\section{Introduction} 
\label{sec:Intro}
Analysis and prediction for economics and finance have been a popular field of study for industrial, governmental, and academic groups \cite{saaty1991prediction}. These studies typically involve large data sets~\cite{wang2010vast} and complicated dependencies with numerous tangible and intangible factors. Prediction in stock markets is especially complex given the volatility of the system. However, the lure of high rewards is motivation enough for many to study these systems extensively. Many works exist in stock-price prediction using statistical models~\cite{islam2020comparison,jothimani2016discrete,atsalakis2009surveying} and time-series analysis~\cite{idrees2019prediction,guo2008time,naik2013does}. 

Prediction of financial markets is often complicated owing to the non-linear dependence of stock prices on several variables~\cite{hommes2001financial,thomaidis2006intelligent}. To circumvent the non-linear dependence of stock prices, a natural solution technique that has been employed is big-data driven machine learning \cite{chong2017deep, heaton2017deep, nikou2019stock, leung2014machine}. While earlier works using traditional machine-learning techniques, such as random forests~\cite{khaidem2016predicting,sharma2017combining}, support vector regression~\cite{henrique2018stock,lee2009using}, shallow Neural networks~\cite{orimoloye2020comparing,singh2017stock}, established proof-of-concept applicability, the works that employ more complex ideas of deep learning, such as LSTMs and encoder-decoder structures are more promising, owing to the time-series nature of the data, i.e. market prices and movements.

In this paper, we present an exploration of candidate architectures motivated by \href{https://www.analyticsvidhya.com/blog/2021/05/stock-price-prediction-and-forecasting-using-stacked-lstm}{this article}~\cite{LSTMstack}, to make stock price predictions. We demonstrate the applicability of our model across several stocks spanning different industry types by using the target architecture obtained from the previous step. To incorporate the effect of unquantifiable events, such as the change in government or company policies, elections, reports of scientific breakthroughs, scandals, etc., we hope to use stock prices in allied markets and industries as proxies. The development of our project can be found here: \href{https://gitlab.com/cs230_stock_analysis/stock_price_prediction}{StockBot}.

The paper is structured as follows. First, we discuss the data acquisition and preparation techniques, followed by the initial exploration of candidate network architectures in Section~\ref{sec:data_methods}. Second, we present our numerical tests in Section~\ref{sec:results}. We present a generalizable price predicting model that can be used to predict stock prices for new stocks, that do not have sufficient historical data. We then present the outcomes of decisions made by our StockBot, which makes stock exchange decisions based on the model's prediction. We conclude our discussion and briefly discuss the next steps to be taken toward a more reliable stock prediction model in Section~\ref{sec:conclusions}.
\section{Data and Methods} \label{sec:data_methods}
\subsection{Data Preparation}
The data used in this study is obtained from the financial index data available on \href{https://finance.yahoo.com}{Yahoo! Finance} \cite{Yahoofin}. We restrict our analysis to stock listed in the NYSE for this study, even though \href{https://finance.yahoo.com}{Yahoo! Finance} \cite{Yahoofin} provides historic data for stocks, ETFs and market indices across the globe. The model is trained on (to infer) the adjusted closing prices of the stock tickers. The dataset is prepared by a moving window that captures stock values for the desired number of days in the past we want to look at (\texttt{past\_history}) against an output vector of length equal to the number of days in the future for which we want to predict stock prices (\texttt{forward\_look}). Even though most results are shown for \texttt{forward\_look}$=1$, our framework is designed to generalize to predicting any number of days in the future in one go. For our representative test cases, we are looking at historic stock data from 2010 to 2020. With a \texttt{train-dev} \texttt{split} of 80$\%$/20$\%$, the training set is comprised of historic data approximately till the end of 2017, while the remaining data is used to create the development set. 

\subsection{Methodology}
To leverage the sequential nature of the financial data, Long Short-Term Memory (LSTM) architectures have been used for price forecasting, which help mitigate the vanishing- and exploding-gradients problems that arise when sequential data is trained using RNNs. We experiment with a number of architectures that include the stacking of LSTM layers of different sizes. We also experiment with \texttt{past\_history} days we look at along with carrying out limited hyperparameter tuning to develop the final model that we use in the discussion in Section~\ref{sec:result}. Additionally, we implement an encoder-decoder model, where we encode historic stock-price data to make future price predictions. Details of each model are discussed in Section~\ref{sec:result}, to compare its performance against the stacked LSTM model.
\subsection{Preliminary Results: Architectures Explored} \label{sec:result}
In this section, we summarize the results from the two architectures we explored. 
We use the Tensorflow Keras module in Python to implement the network. We choose the mean-squared error loss function and Adam optimization for training the weights.
The default values of the hyperparameters we use are shown in Table.~\ref{tab:my_label} (can also be accessed from our source code available on GitLab): 
\begin{table}[H]
    \centering
    \begin{tabular}{|l|c||l|c|}
    \hline
        Hyperparameter & Value & Hyperparameter & Value \\
        \hline
         \texttt{past\_history} & 60 & Batch size & 64 \\
         LSTM layer units & 20 & Epochs & 500 \\
         Stack depth & 2 & Steps per epoch & 200 \\
         Train-test split & 80\%/20\% & Validation steps & 50 \\
         \hline
         \end{tabular}
    \caption{Hyperparameters used for optimal test-set performance}
    \label{tab:my_label}
\end{table}
 
\subsubsection{LSTM based models}
\begin{figure}[!htb]
    \centering
    \subfigure[]{\includegraphics[width=0.3\textwidth]{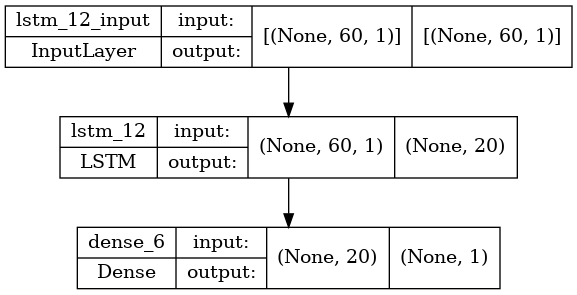}}
    \subfigure[]{\includegraphics[width=0.215\textwidth]{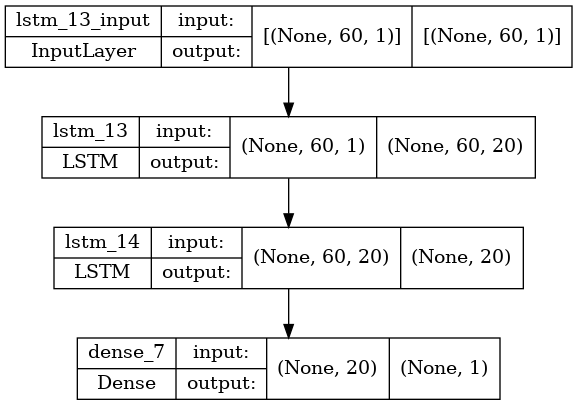}}
    \subfigure[]{\includegraphics[width=0.38\textwidth]{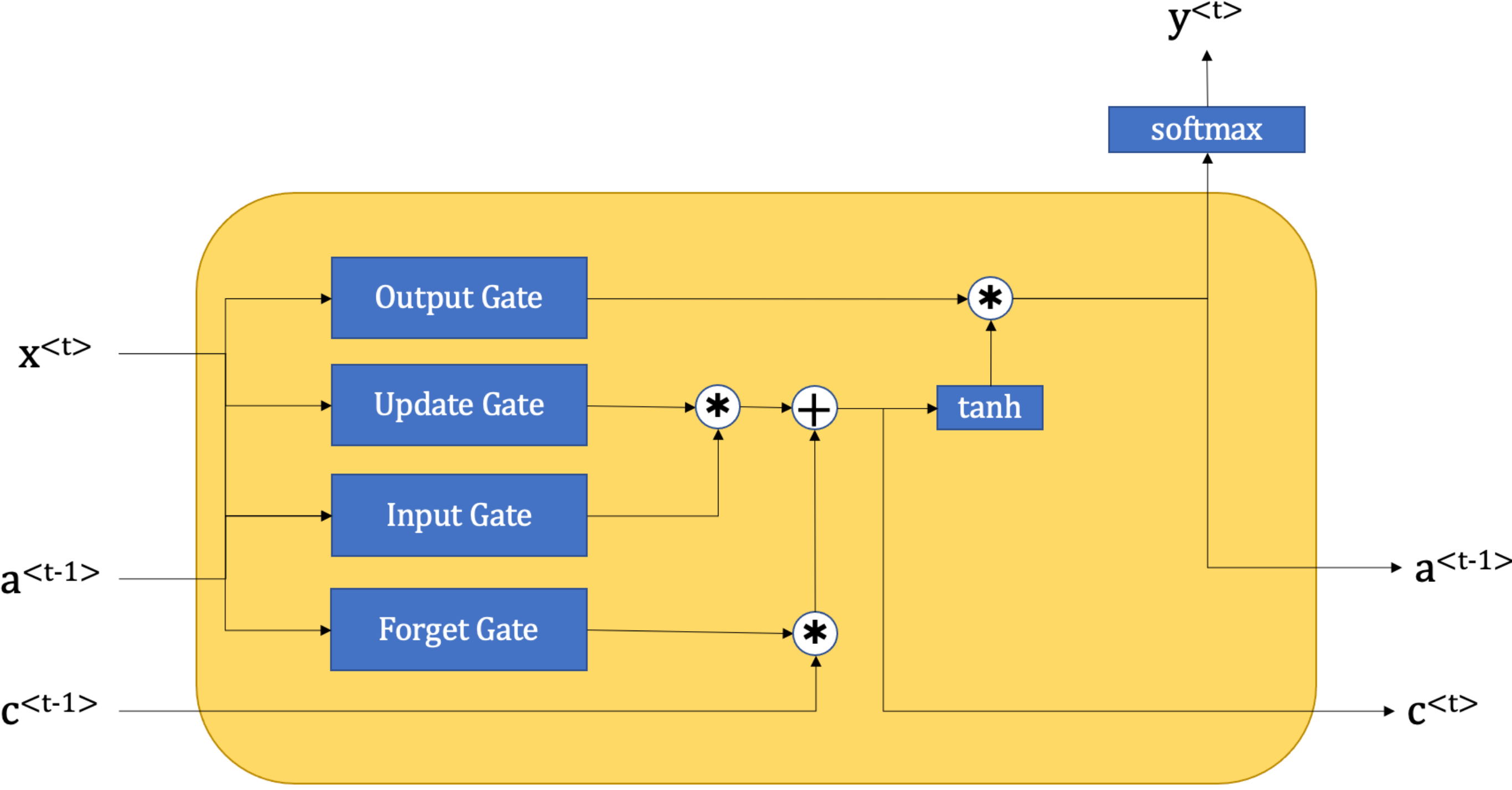}}
    
    \subfigure[]{\includegraphics[width=0.98\textwidth]{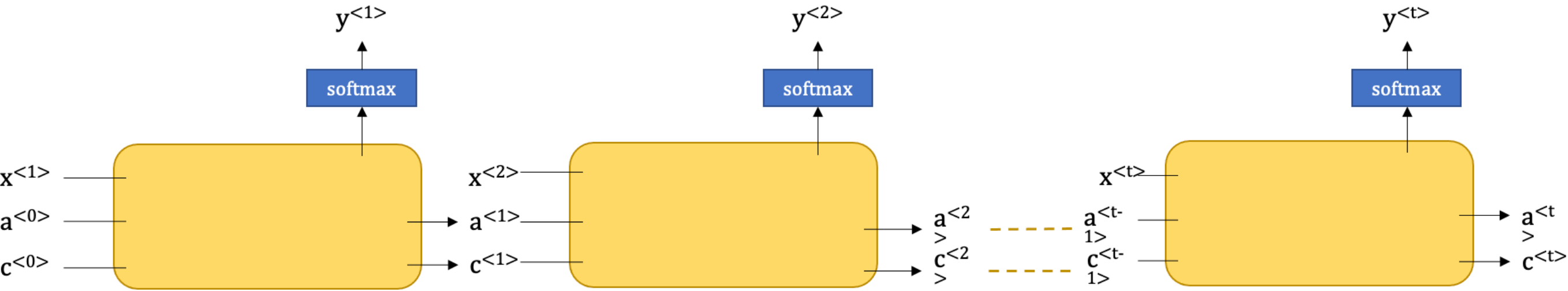}}
    \caption{Block diagrams representing architectures for (a) single layer LSTM and (b) double layer LSTM. (c) Typical LSTM cell and (d) arrange LSTM cells as part of the network. \label{fig:LSTM_arch}}
\end{figure}

We discuss our experiment on single/stacked many-to-one LSTM architectures in this section (Fig.~\ref{fig:LSTM_arch}).

The compact form of the equations within a block while executing the forward pass are summarized by Eqs.(\ref{eq:first}-\ref{eq:last}).
\begin{align}
    f_t&=\sigma_g(W_fx_t+U_fh_{t-1}+b_f), \label{eq:first}\\
    i_t&=\sigma_g(W_ix_t+U_ih_{t-1}+b_i),\\
    o_t&=\sigma_g(W_ox_t+U_oh_{t-1}+b_o),\\
    \tilde{c}_t&=\sigma_c(W_cx_t+U_ch_{t-1}+b_c),\\
    c_t&=f_t \odot c_{t-1} + i_t\odot\tilde{c}_t,\\
    h_t&= o_t \odot \sigma_h(c_t) \label{eq:last},
\end{align}
where $x_t$ is the input vector the LSTM unit. $f_t$, $i_t$, and $o_t$ refer to the activation of the forget gate, input gate and output gate respectively. $h_t$ is the hidden state vector, $\tilde{c}_t$ is the cell input state vector and $c_t$ is the cell state vector. $W$, $U$, and $b$ are the weights and biases that are learned while training the LSTM network.

The final results are presented by predicting the future 200 days' closing price prediction, one day at a time from a model trained on looking at the past 60 days' closing stock price. We discuss two methods to predict stock prices. One method is to use the ground-truth closing prices available over a long trajectory in the development set for every instance of prediction. The other method is to use previously-predicted stock prices to predict subsequent values. In practice, we do not have any knowledge of the future stock prices, so our prediction should be based on the second method. We present our results using both methods, i.e., i) ground-truth-based prediction (infeasible in practice, serves as a reference), and ii) dynamically-updated-truth-based prediction (of practical use). We observe that both methods perform comparably, implying that the stock price trajectory predicted by our model is stable in predicting even when all the data the model sees are previously-predicted values and not the ground truth.
\begin{figure}[!htb]
    \centering
    \subfigure[]{\includegraphics[width=0.32\textwidth]{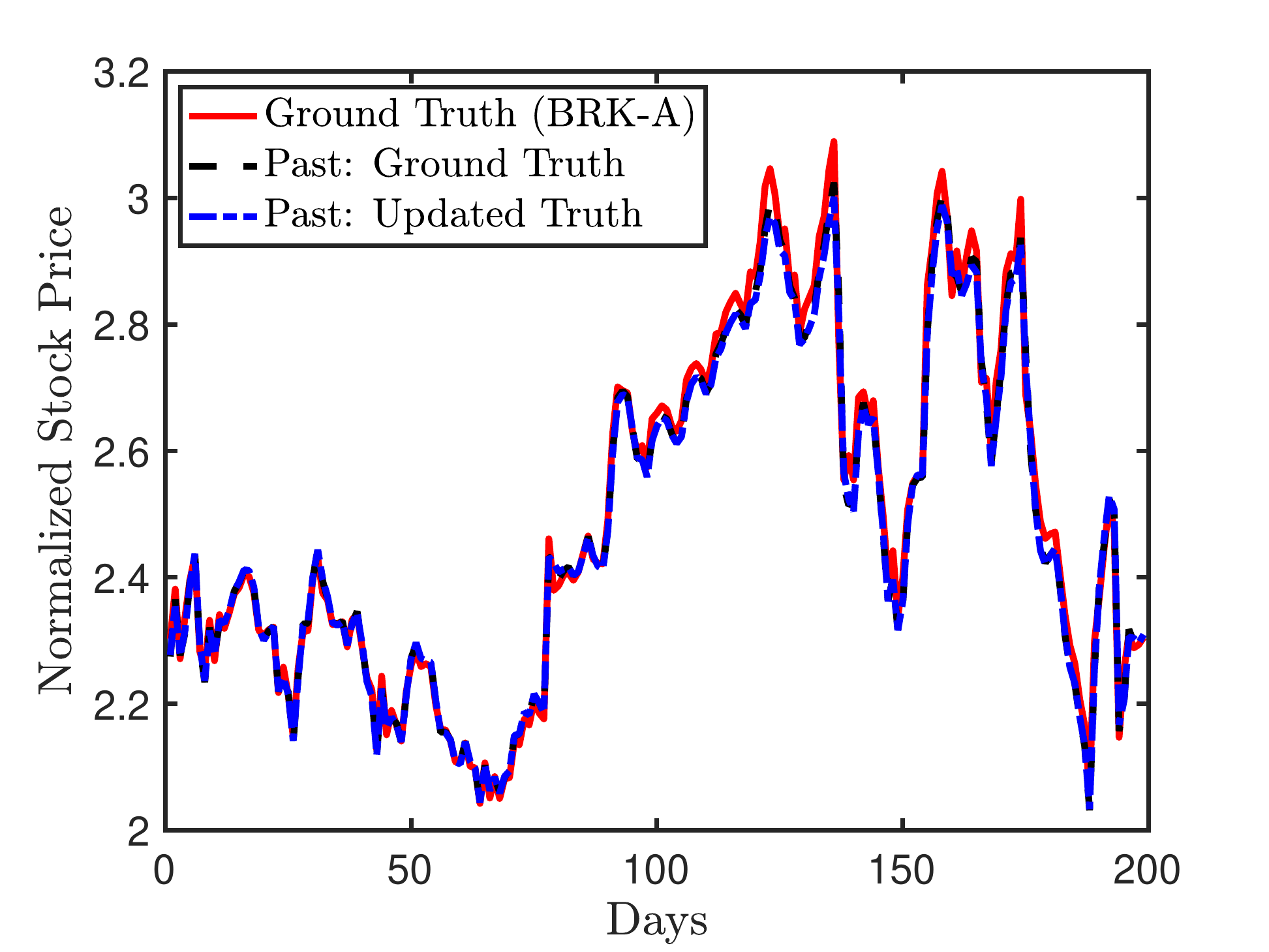}}
    \subfigure[]{\includegraphics[width=0.32\textwidth]{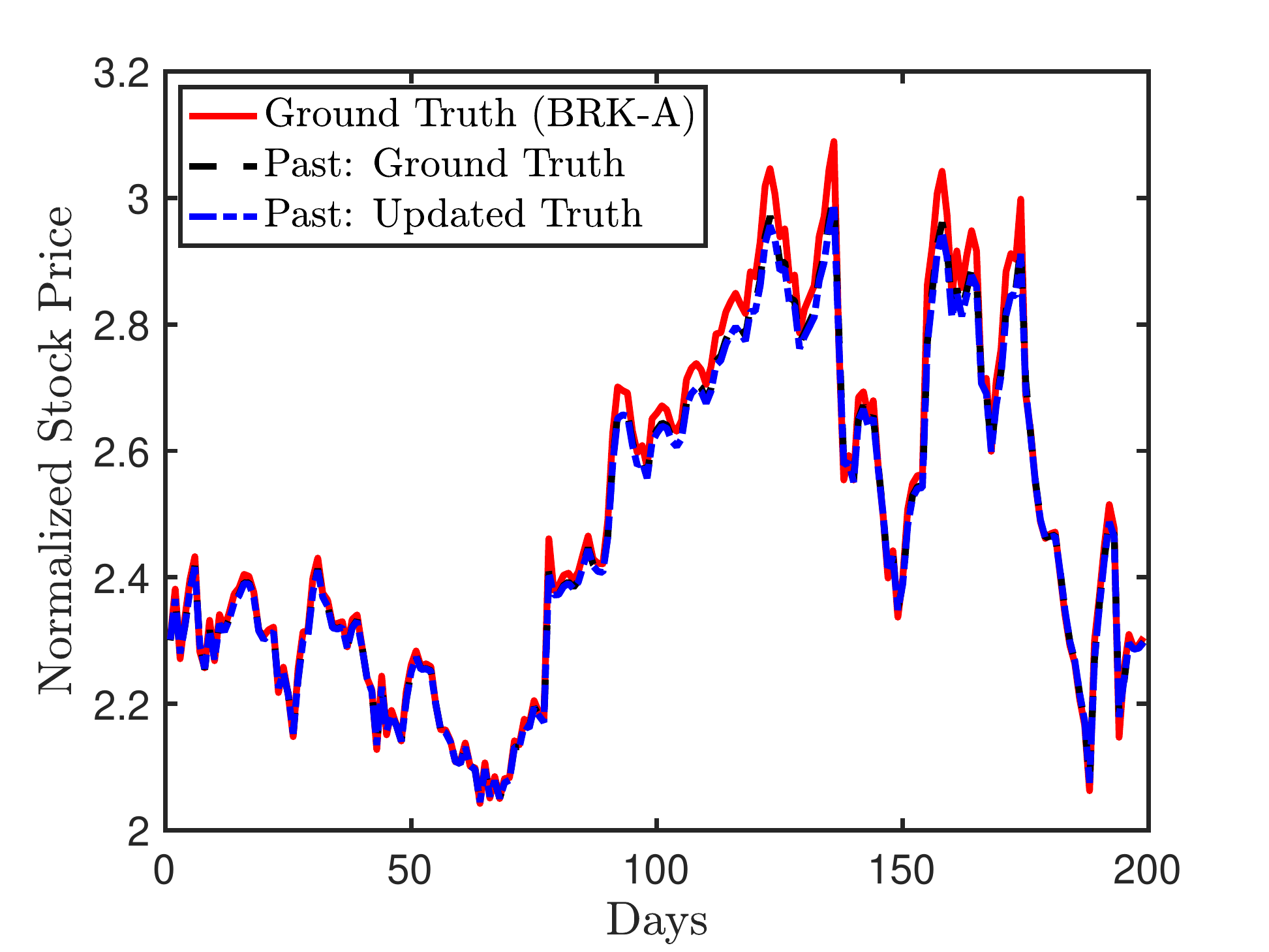}}
    \subfigure[]{\includegraphics[width=0.32\textwidth]{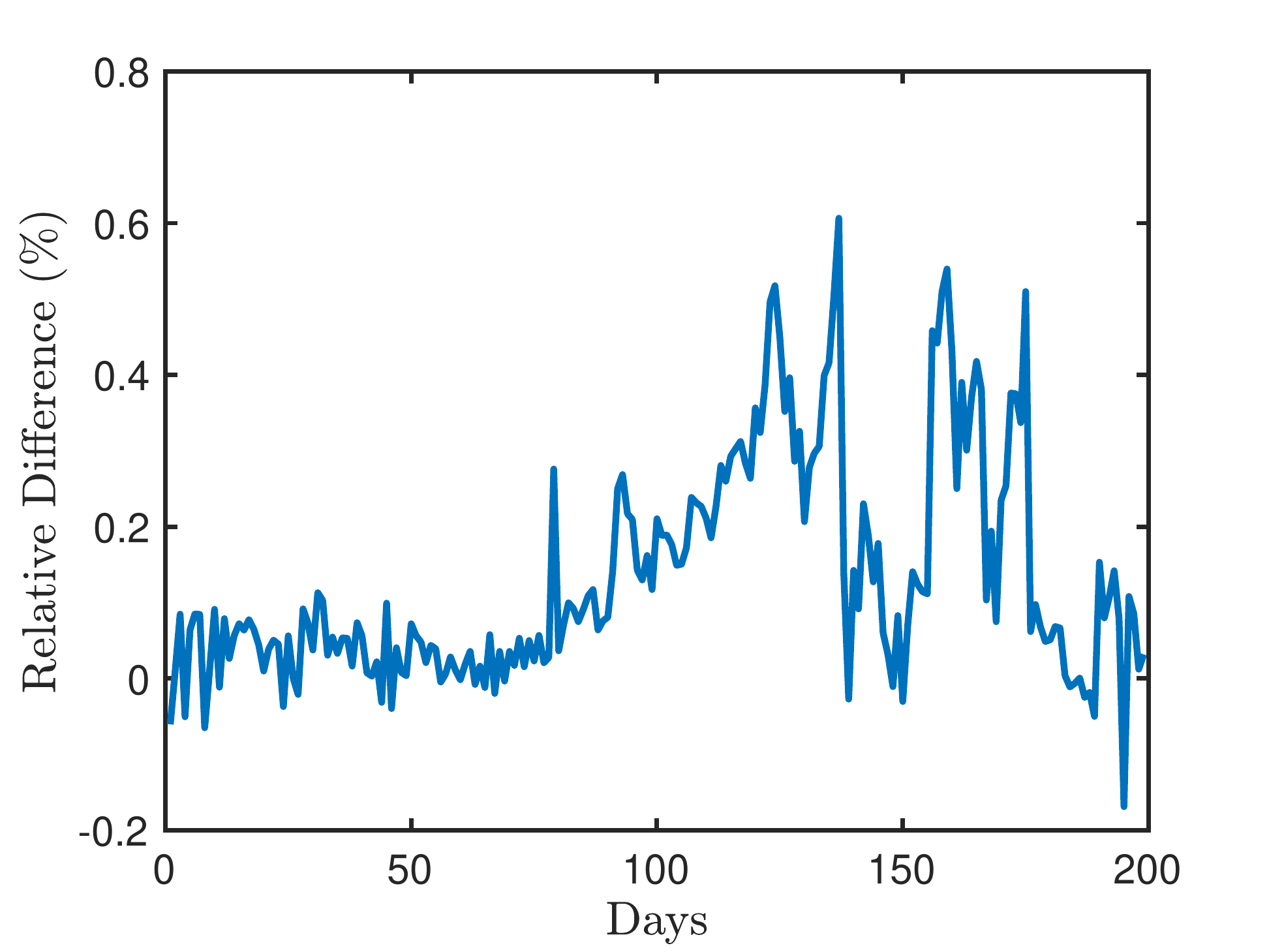}}
    \caption{Predicted values using the ground truth and the updated truth as the history for (a) a single layer LSTM and (b) a double layer LSTM. (c) The difference between the prediction using the ground truth and the updated truth.  \label{fig:LSTM_base}}
\end{figure}

From our preliminary predictions for the stock ticker \texttt{BRK-A} (Berkshire Hathaway), we see that a single LSTM layer outperforms a stacked LSTM network when we exclusively train a model for a given stock, as shown in Fig.~\ref{fig:LSTM_base}. We also see that the model is robust in capturing the price change with time, with the predictions from the updated truth being almost as good as the predictions from the ground truth, as shown in the relative difference in Fig.~\ref{fig:LSTM_base}.

\begin{table}[H]
    \centering
    \begin{tabular}{|c|c|c|c|c|c|c|}
    \hline
     & \multicolumn{2}{|c|}{BRK-A} & \multicolumn{2}{|c|}{GOOG}& \multicolumn{2}{|c|}{MSFT} \\
    \hline
        Method & Train & Test& Train & Test& Train & Test \\
        \hline
       Ground Truth  & 0.0011 & 0.0071 & 0.0011 & 0.0044 & 0.0012&  0.0167\\
       \hline
       Updated Truth  & 0.0035 & 0.0049 & 0.0035& 0.0051 & 0.0054&0.0054 \\
         \hline
    \end{tabular}
    \caption{Single Layer LSTM: The train and test MSE.}
    \label{tab:my_label_1}
\end{table}

From the train and test RMS error, we can see that the updated truth predictions do just as well as the ground truth predictions, as shown in Tables. ~\ref{tab:my_label_1} and \ref{tab:my_label_2}. We also establish the stability of the model in capturing the stock price over the next 200 days.

\begin{table}[H]
    \centering
    \begin{tabular}{|c|c|c|c|c|c|c|}
    \hline
     & \multicolumn{2}{|c|}{BRK-A} & \multicolumn{2}{|c|}{GOOG}& \multicolumn{2}{|c|}{MSFT} \\ 
    \hline
        Method & Train & Test & Train & Test & Train & Test \\
        \hline
       Ground Truth  & 0.0011 & 0.0063 & 0.0012 & 0.0080& 0.0011 & 0.022\\
       \hline
       Updated Truth  & 0.0036 & 0.0049 &0.0055 & 0.0056 & 0.0050&0.0044 \\
         \hline
    \end{tabular}
    \caption{Double Layer LSTM: The train and test MSE.}
    \label{tab:my_label_2}
\end{table}
However, we see that the Microsoft (`MSFT') stock prediction does not agree well with the actual stock closing prices (Figure not part of the report to preserve brevity), even though it successfully reproduces the qualitative trend which should be sufficient to make buy/sell decisions, discussed later in Section~\ref{sec:buy_sell_bot}. To mitigate the quantitative error observed during inference, we could increase the size of our training data by considering more historic data, but we want to explore the data from tickers from a similar industry to be able to reduce the bias observed in the training set prediction. This multi-stock model is discussed later in Section~\ref{sec:multi_stock}
\subsubsection{Encoder-Decoder Model}
In this section, we will describe the Encoder-Decoder architecture that we are currently experimenting with. The model is influenced by the character-level English to French translation algorithm described in this article \cite{EDref}. It is worth noting that there is a parallel that can be drawn between character-level translation and stock price prediction. Holding a price history array analogous to a sentence in English, a sentence of ``future prices'' can be obtained by ``character-wise'' translation, the characters being analogous to single price entries.

\begin{figure}[!htb]
    \centering
    \subfigure[]{\includegraphics[width=0.65\textwidth]{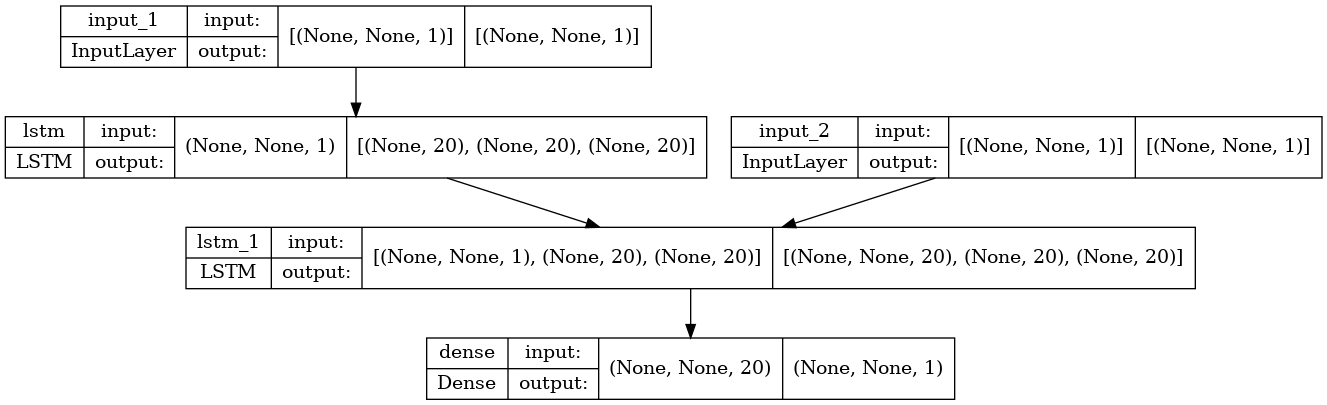}}
    \subfigure[]{\includegraphics[width=0.32\textwidth]{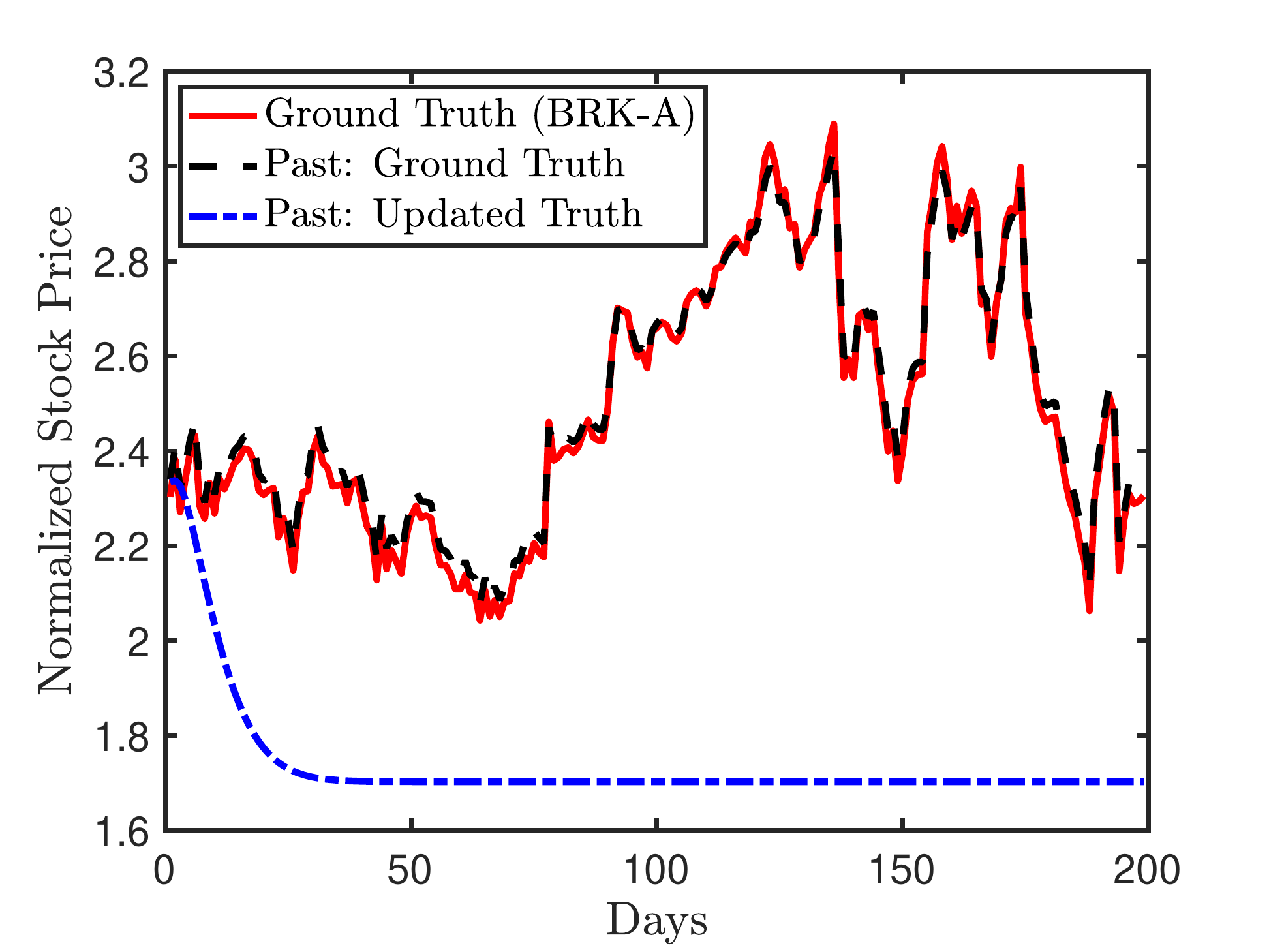}}
    \caption{(a) Block diagram representing Encoder-Decoder architecture. (b) Predicted values using the ground truth and the updated truth as the history for Encoder-Decoder model}
    \label{fig:LSTM_ED_arch}
\end{figure}
The model consists of two LSTM blocks, `lstm' as encoder and `\texttt{lstm\_1}' as decoder as shown in Fig.~\ref{fig:LSTM_ED_arch}. The input to the encoder is past 60 days' closing stock price. During training, in addition to receiving the encoder state, the decoder receives an input of ground-truth future values up to one day before the last day for which price is to be predicted (``teacher forcing''). Using these inputs the decoder is trained to output information related to future predictions (up to the required number of days), which is then passed to a dense layer that outputs the final predicted prices. During prediction, however, the decoder should function using its predicted values at each step. In the current configuration, the model takes in 60 days' price history and outputs the price prediction for the 61st day alone.

Similar to the LSTM model above, the Encoder-Decoder model was used to predict future stock prices using ground truth values and self-updated values separately. The results of this are seen in Fig.~\ref{fig:LSTM_ED_arch}. We see that the model is currently not making satisfactory predictions when it predicts dynamically, as opposed to when receiving ground truth future values to give predictions offset by one day. A possible reason for this is the amplification of prediction error at each time step. To mitigate this issue, future work can be focused on enhancing the model by making the architecture more complex so that this fundamentally translation-motivated model becomes better optimized for stock prediction.

\section{Results} \label{sec:results}
We see that our single stack and double stack LSTMs perform better than most of the more computationally intensive network architectures enlisted~\cite{paperswithcode}. In order to address common issues with stock prediction architectures such as: (i) capture a more robust prediction for stocks without sufficient historical data, (ii) improve prediction inference time (in the case of second-to-second updates which are in practice unlikely to be used), (iii) extracting investment insights, we present a set of numerical experiments in this section.

\subsection{Multi Stock Model} \label{sec:multi_stock}
So far, all the models we have discussed involve treating each stock separately. i.e., the stock price corresponding to each ticker is fed into the model one at a time, and predictions are made for this stock using the parameters trained on the same stock. Given sufficient resources, individual models could be trained but such a model doesn't capture the ticker development across companies of the same industry. After having identified the single/double-stacked LSTMs as our target network architecture, we present a more generalizable model that is trained specifically to an industry type such as ``energy'' or ``finance'', by accounting for the historic data of all prominent tickers within that industry.

\begin{figure}[!htb]
    \centering
    \subfigure[]{\includegraphics[width=0.31\textwidth]{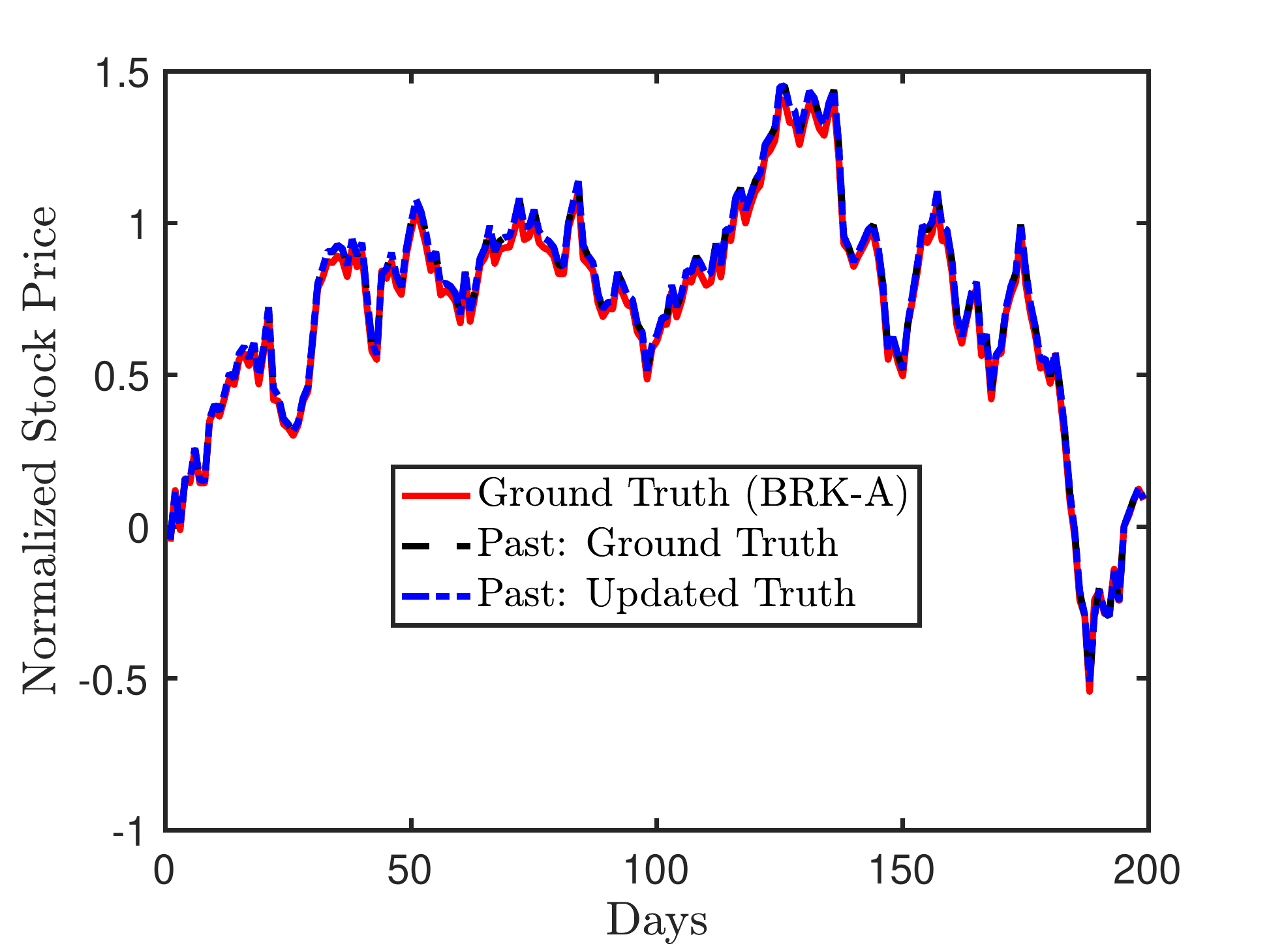}\label{fig:ms_pred} } 
    \subfigure[]{\includegraphics[width=0.31\textwidth]{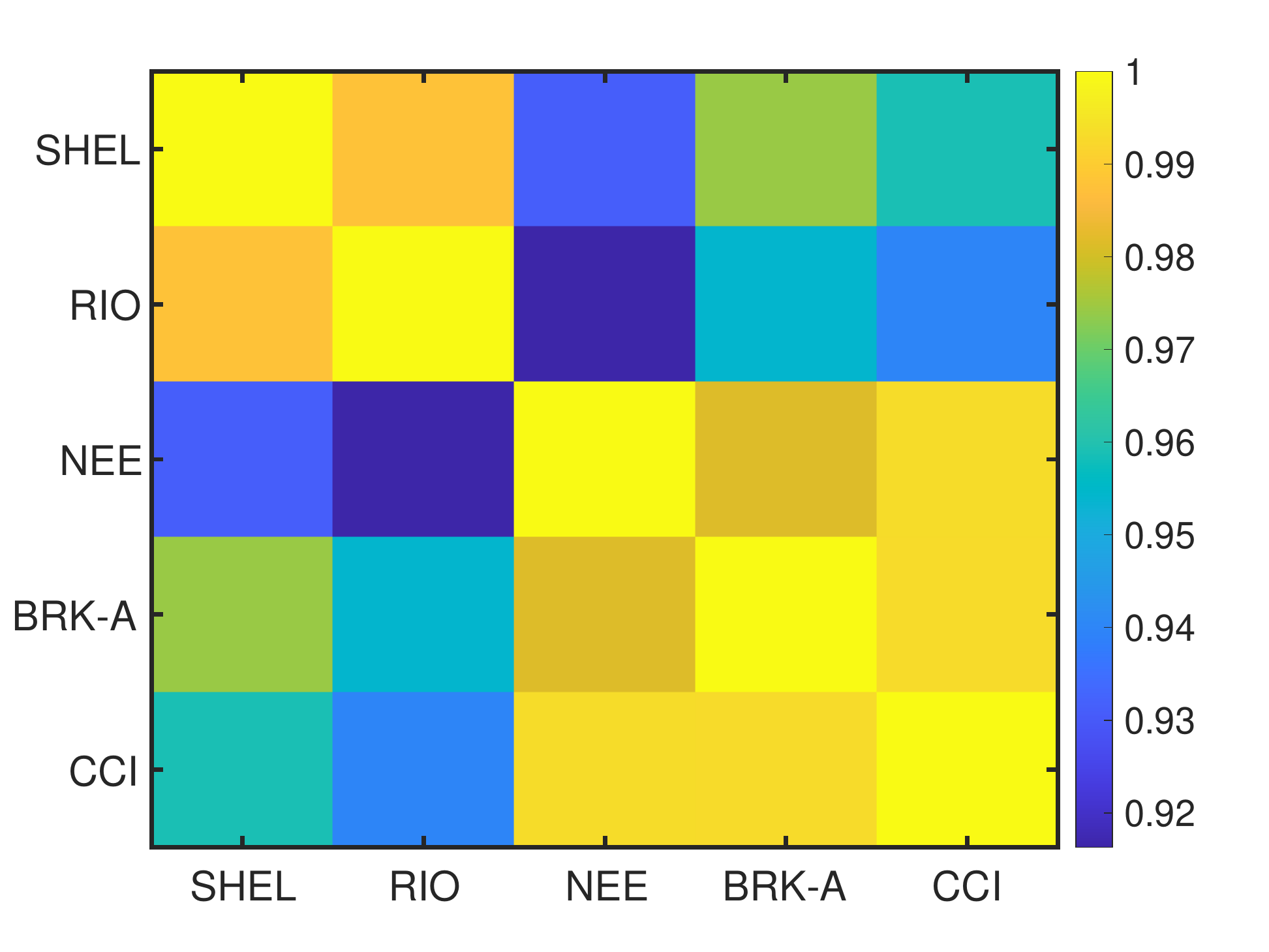} \label{fig:cc_cc}} 
    \subfigure[]{\includegraphics[width=0.31\textwidth]{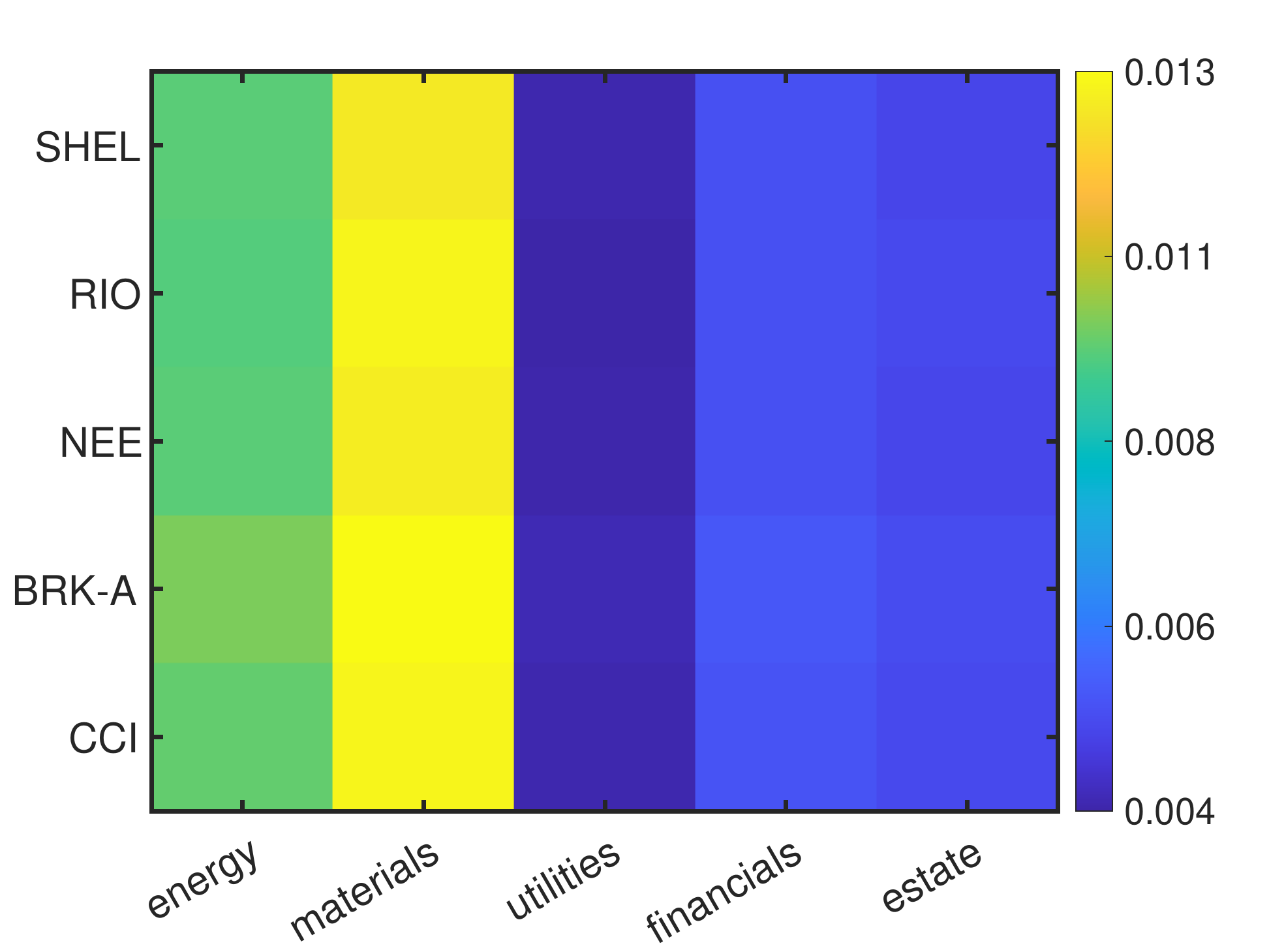} \label{fig:cc_rms}} 
    \caption{(a) Predicted values of the `XOM' ticker using the ground truth and the updated truth as the history for a stacked LSTM (model trained on the historic data of other energy industry stocks). (b) The correlation between five stock tickers (`SHEL', `RIO', `NEE', `BRK-A', `CCI') which belong to five different industry categories on the NYSE. (c) The corresponding test RMS error of predicting given stock value by using models trained across different industry stocks.  \label{fig:LSTM_correl}}
\end{figure}

The ``Multi Stock'' model is essentially the same single/stacked LSTM model but trained on data prepared in a different manner. It involves taking several tickers from a given industry and creating a mixed training and/or test set which involves past price - future price combinations of all these tickers. The model is defined to function in two modes. When we set `\textit{sameTickerTestTrain} = True' both train and test sets include sets of data for multiple tickers. For `\texttt{sameTickerTestTrain = False}',  a single ticker is isolated and kept aside and past price - future price combinations for the rest of the tickers are fed into the training set. The dev and test sets are fed with data from the isolated ticker. The latter mode gives a better idea of how well we can estimate the prices of a relatively new stock in a pre-existing industry, as shown in Fig.~\ref{fig:ms_pred}.

Furthermore, we look at models trained on five different industries by arbitrarily removing one ticker from each to test on. We try forecasting the stock prices by all five models, where we hypothesize that the weakly correlated stocks (shown in Fig.~\ref{fig:cc_cc}) would perform the worst while inferred from each other's models. This hypothesis is captured on our RMSE grid, as shown in Fig.~\ref{fig:cc_rms}.

\subsection{Predicting multiple days together}
\begin{figure}[!htb]
    \centering
    \subfigure[]{\includegraphics[width=0.32\textwidth]{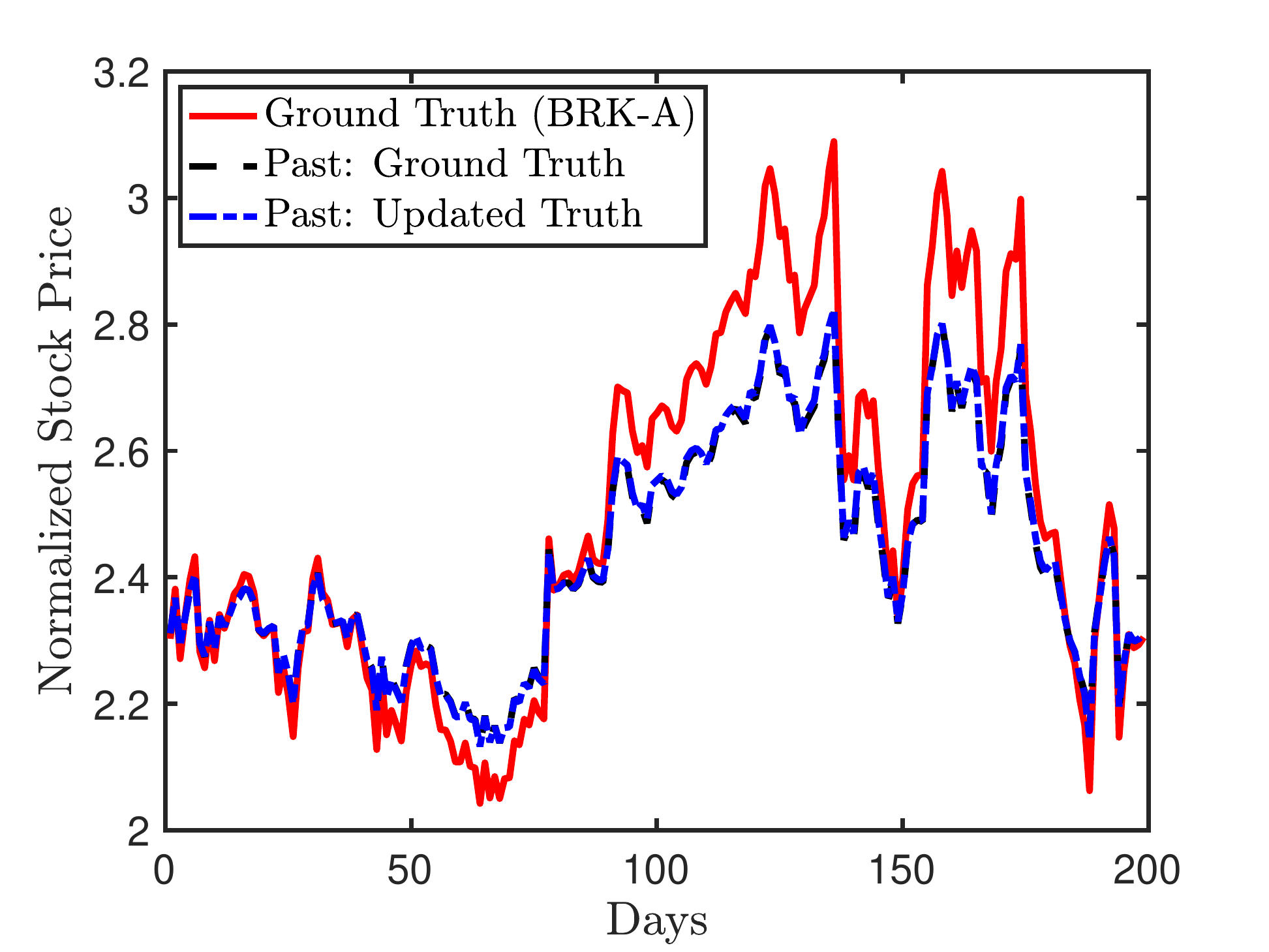} \label{fig:pre_weight}}
    \subfigure[]{\includegraphics[width=0.32\textwidth]{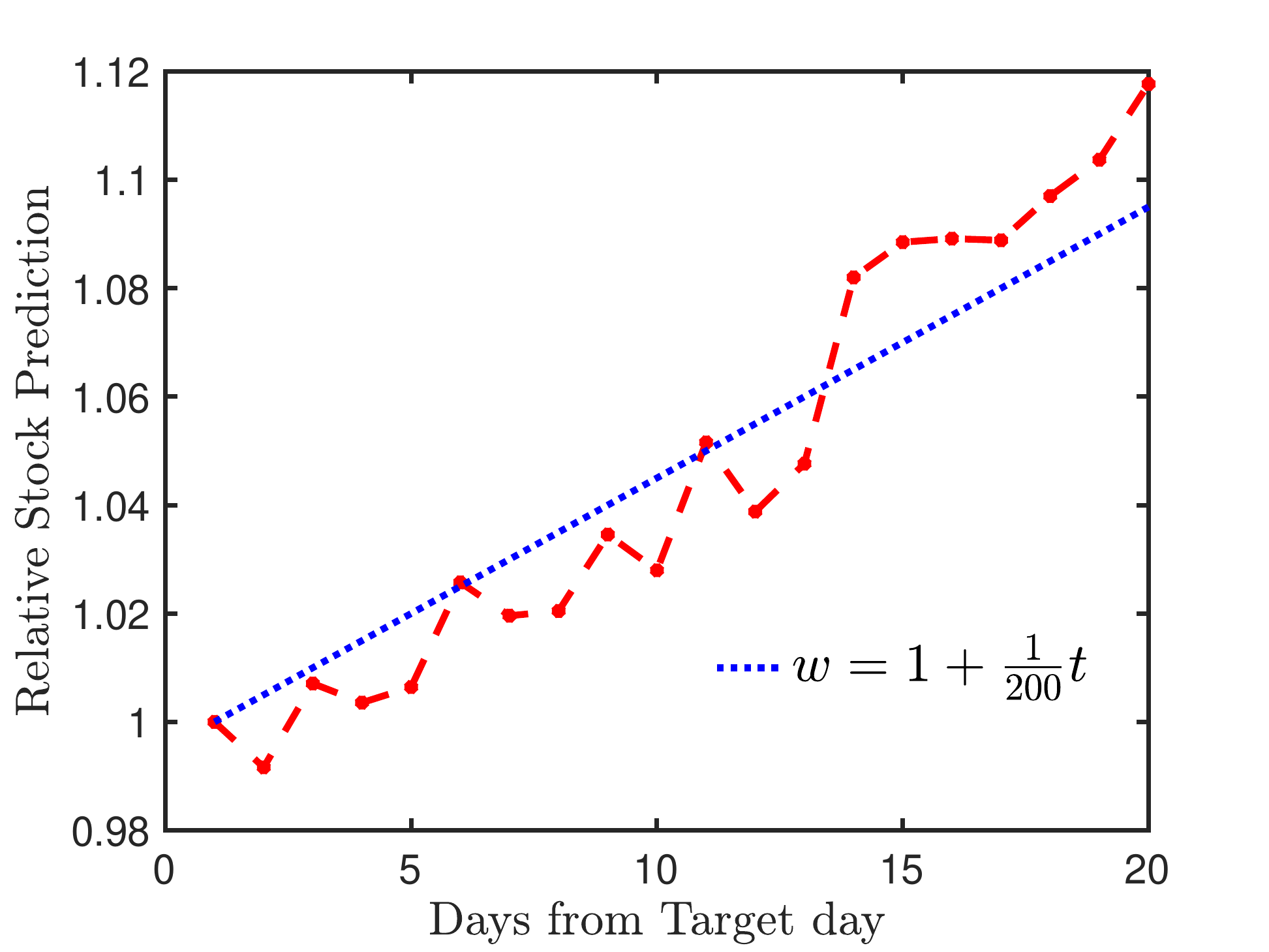}\label{fig:weight}}
    \subfigure[]{\includegraphics[width=0.32\textwidth]{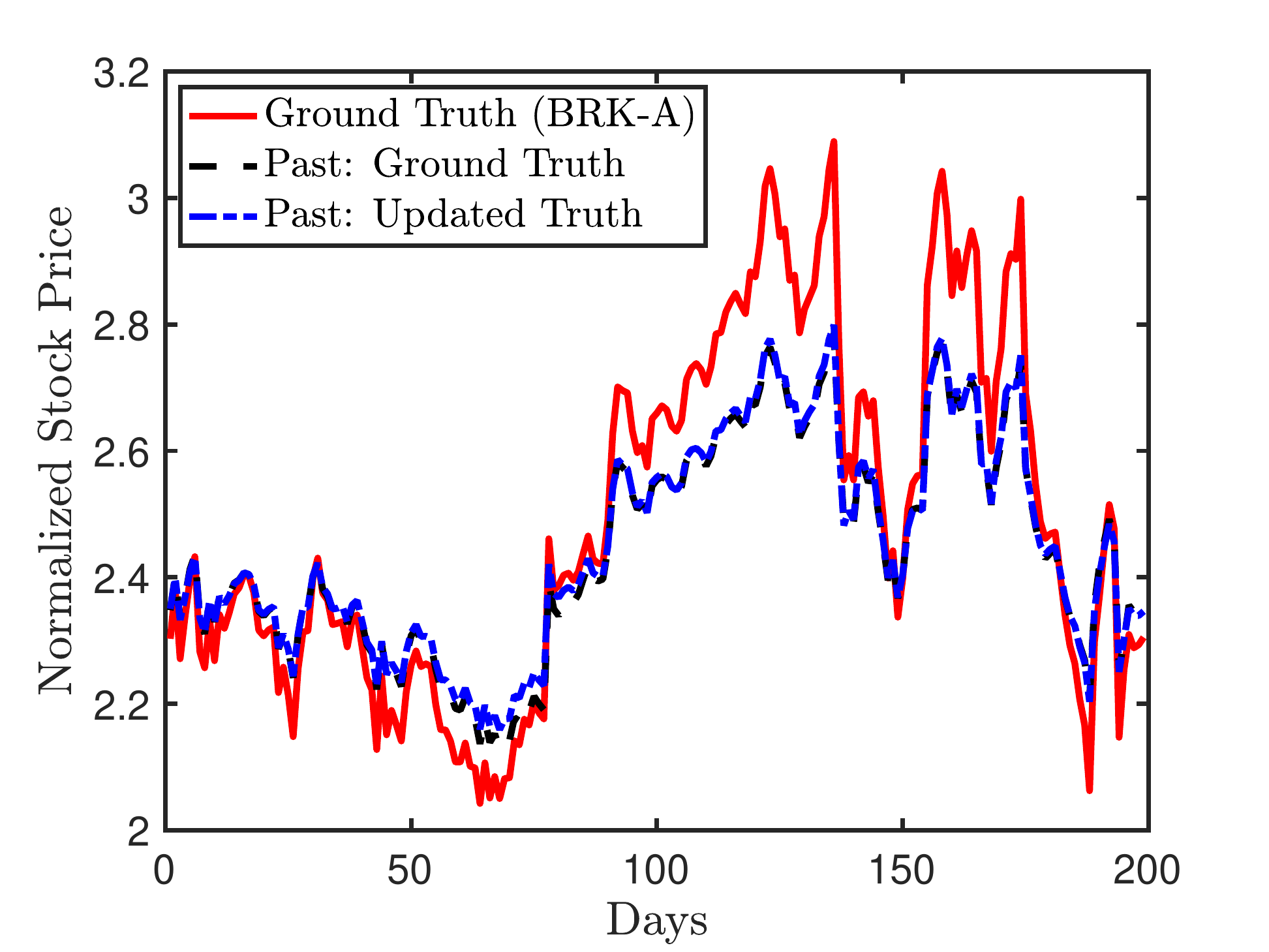}\label{fig:post_weight}}
    \caption{Predicted values of the `BRK-A' by predicting 20 days ahead a a time (updating the ground truth one day at a time) using the (a) mean squared error loss and the (c) weight adjusted mean squared error loss. (b) The convergence of the stock predicted values as a function of the number of days we are away from the target prediction day. 
    \label{fig:LSTM_future_fit}}
\end{figure}

We see that a sequential update of the dataset to predict the stock price in the future is almost as good as predicting the future day one at a time, by using the actual data. However, this method makes inference a slower process by limiting the prediction to just one day in the future at a time. Predicting multiple days in the future may come with its shortcoming where the prediction of the stock for day $t+n$ (while predicting $[t+1,\, t+n]$) by looking at $[t-k,\, t]$, turns out to be much worse than predicting for the same day while predicting the sequence $[t+n,\, t+2n-1]$ by using $[t+n-1-k,\, t+n-1]$. We continue updating the ground truth one day at a time, while continuing to predict for multiple days (20 days) in the future instead of updating the ground truth for all the predicted days since even the single day update of the ground truth performs significantly worse, as shown in Fig.~\ref{fig:pre_weight}. A deeper or a larger model will need to be trained for accurately predicting multiple days ahead at a time which can make the inference $~10^2$-$10^3$ times longer. This shows that i) training a small model to predict just one day in the future is much more accurate and ii) inferring stock prices from a model designed to give accurate stock predictions is much more computationally efficient when designed for predicting one day in the future at a time. 

Interestingly, we see that the prediction converges as we move closer to the target prediction day, as shown in Fig.~\ref{fig:weight}. For the sake of simplicity, we fit a line through the prediction convergence to obtain the correction weights, 
\begin{equation}
    w\approx 1+\frac{1}{200}t
\end{equation} 
for the loss function. Here $t$ is the number of days we are away from the target prediction day. We now modify the original loss function 
\begin{equation}
    \frac{1}{m}\sum_i \|\hat{y}^i - y^i \|^2,
\end{equation}
such that it penalizes (with a weight vector $\vec{w}$) the prediction made on the days further ahead, give by,
\begin{equation}
    \frac{1}{m}\sum_i \|\vec{w}\cdot(\hat{y}^i - y^i )\|^2.
\end{equation}Note that the prediction, in this case, $\hat{y}^i$ is a vector of shape (\texttt{forward\_look} ,) where \texttt{forward\_look} $=20$ for the presented results. We see that the predictions obtained after applying the modified loss function are more in agreement with the ground truth, as shown in Fig.~\ref{fig:post_weight}. The most encouraging result of the modified loss function is that the test loss decreases by over 30$\%$ (from 0.091 to 0.059).

\subsection{Stock predicting bot}\label{sec:buy_sell_bot}
In this section, we present the performance of a bot that makes buy/sell operations at the time of closing every day, to maximize gains. The bot makes decisions based on our predictions of the stock prices. These decisions are analytically enforced and are not learned separately. The bot decision is made by noting a two-step process: We first document the $\delta_i$ changes which are defined by $\delta_i = \text{sign}(c_{i+1} - c_i)$, where $c_i$ is the stock price on the $i^{\rm th}$ day. We then look at the changes in the $\delta_i$, by tracking $\Delta_i = \delta_{i+1}-\delta_i$. $\Delta = -2$ corresponds to the end of a dip, where we buy whereas $\Delta_i=2$ corresponds to the beginning of a dip where we sell. The algorithm is summarized below:
\begin{algorithm}[H]
Obtain the predicted trajectory, $\bm{c}$ (stock forecast for \texttt{forward\_look} days;

Compute the nature of change: $\delta_i = \text{sign}(c_{i+1} - c_i)$ ; 

Compute the curvature of the forecast: $\Delta_i = \delta_{i+1}-\delta_i$;

Make the decision: $\text{Decision} = \begin{cases} \Delta_i = 2 \rightarrow \text{ sell (Indicates local maxima)} \\  \Delta_i = 0 \rightarrow \text{ hold (Indicates no change)} \\ \Delta_i = -2 \rightarrow \text{ buy (Indicates local minima)}. \end{cases}$

 \caption{: StockBot decision making algorithm}
\end{algorithm}

\begin{figure}[!htb]
    \centering
    \subfigure[]{\includegraphics[width=0.32\textwidth]{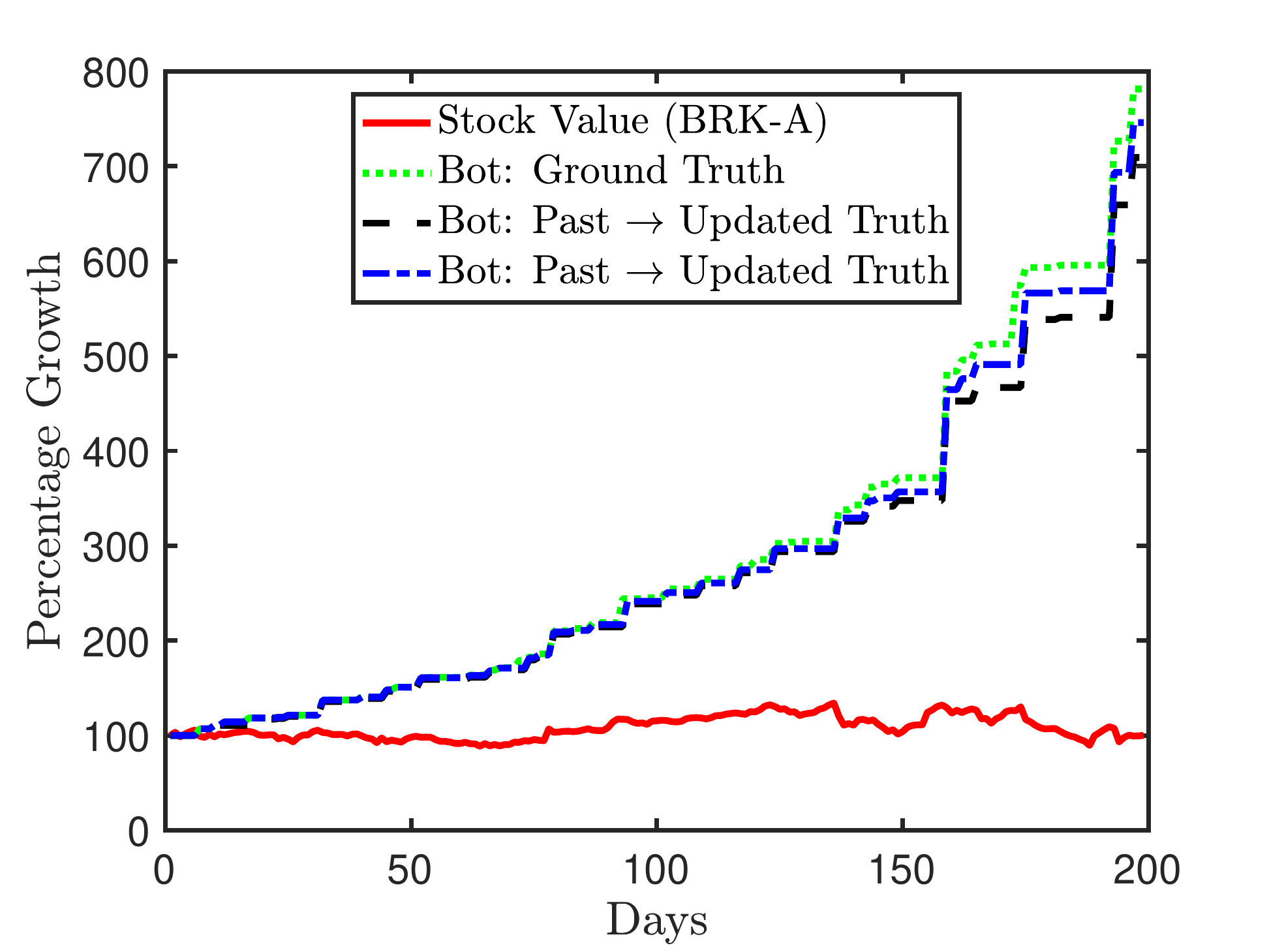}}
    \subfigure[]{\includegraphics[width=0.32\textwidth]{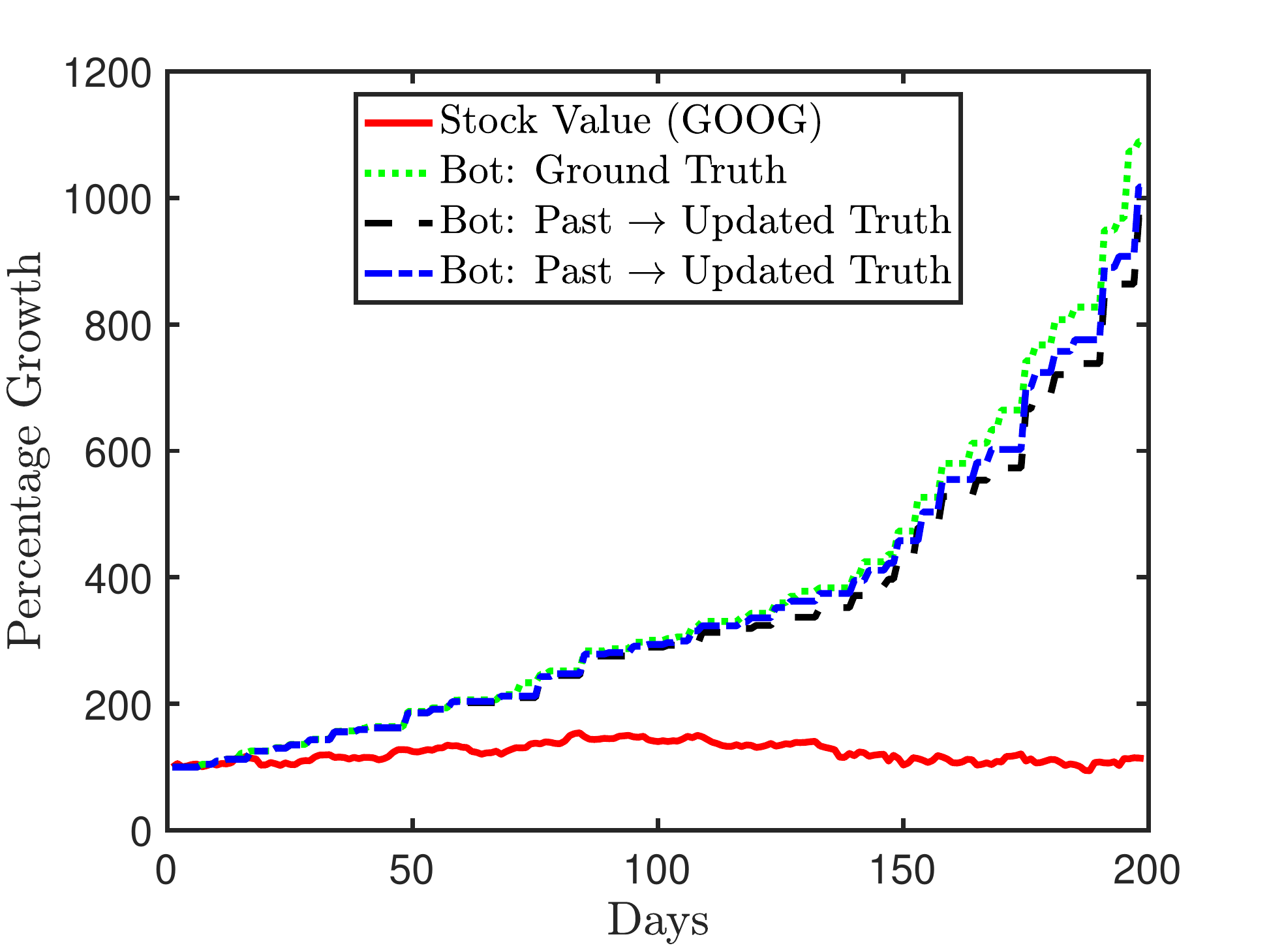}}
    \subfigure[]{\includegraphics[width=0.32\textwidth]{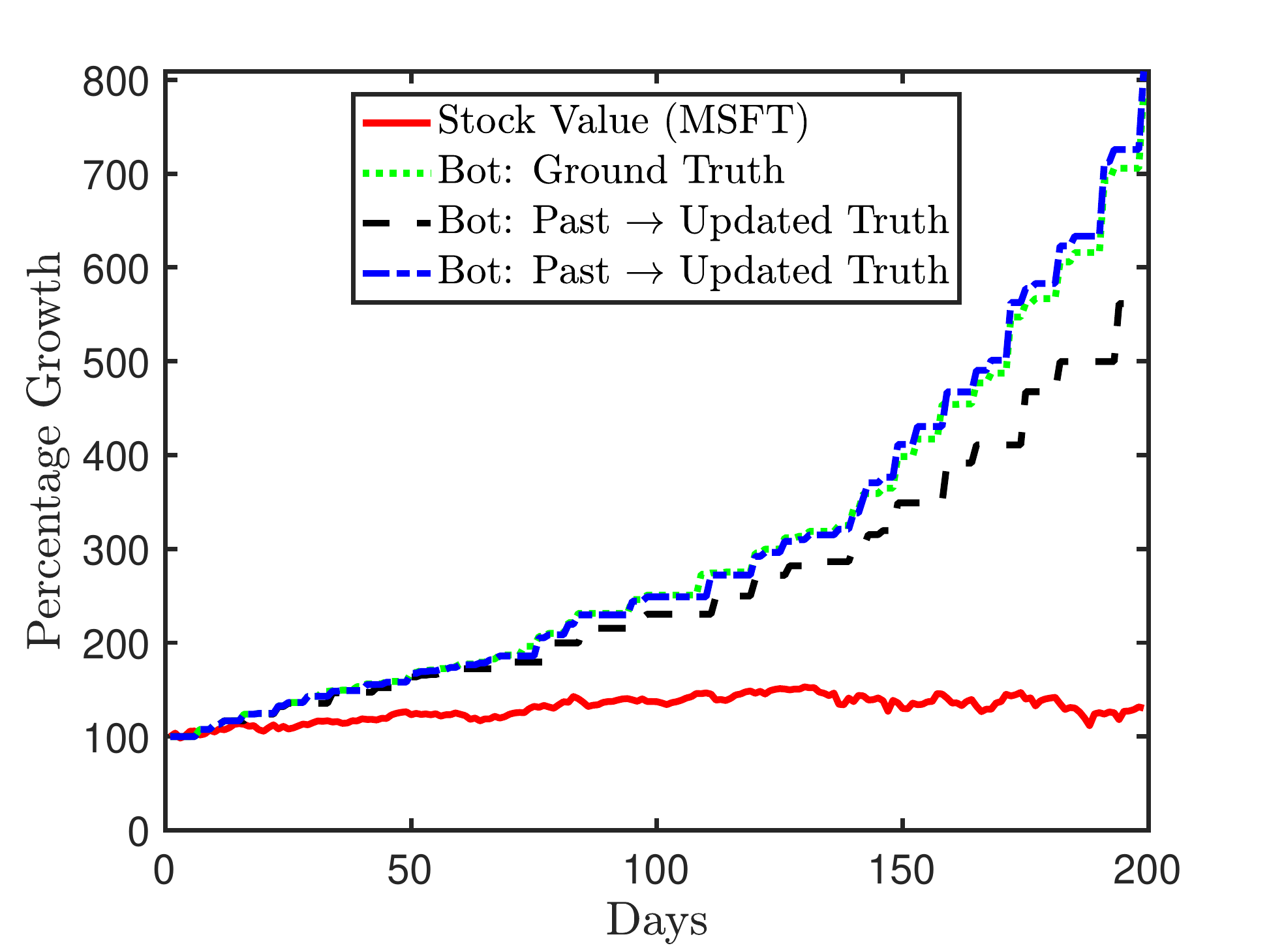}}
    \caption{Percentage growth of invested amount by the bot using a double-stacked LSTM on the (a) BRK-A, (b) GOOG and (c) MSFT tickers. \label{fig:Bot_run}}
\end{figure}

Using this decision strategy on the basis of our future stock price prediction, we pre-decide the buy/sell decisions for $200$ days in advance for our results shown in Fig.~\ref{fig:Bot_run}.

The performance of our bot significantly outperforms even the most aggressive ETFs, as well as leading investment products provided by investment firms. The typical 5-year growth for the best performing ETFs is $<125\%$, which highlights the performance of our bot which yields returns that are over five times greater in period of 200 trading days (equivalent to 10 months). To avoid day trading, the buy and sell frequency could be reduced to once  a week instead of once a day, at the cost of reduced profits. Our bot would yield significant profits even when the stock price itself doesn't exhibit net growth. It should be noted that, for a successfully performing bot, the model should capture the qualitative trends accurately. Our model is not trained for adversarial examples corresponding to unprecedented market crashes but for a fair-weather market, 

\subsection{Google Trends data as additional input}
A possible proxy for other intangible factors that affect the stock prices is regional internet search patterns related to specific relevant keywords. Preis et al. \cite{gtrends2013pricefall} detected an increase in Google search volumes for keywords relating to financial markets, preceding market falls. Motivated by this, we have included a separate model that uses Google Trends data for ticker-related keywords as additional inputs. Google Trends Anchor Bank \cite{GTAB} has been used to obtain unnormalized search volume history related to keywords of the form `\texttt{company\_name} stock'. Past history of the search volume data is fed as an extra dimension parallel to stock price history. 

\begin{figure}[!htb]
    \centering
    \subfigure[]{\includegraphics[width=0.45\textwidth]{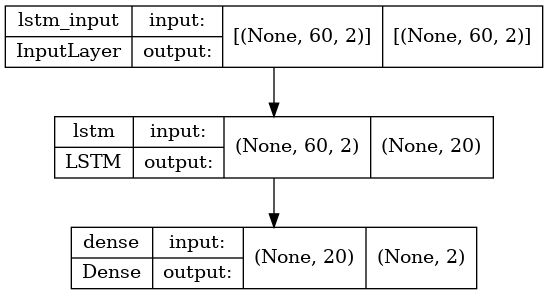} }
    \includegraphics[width=0.4\textwidth]{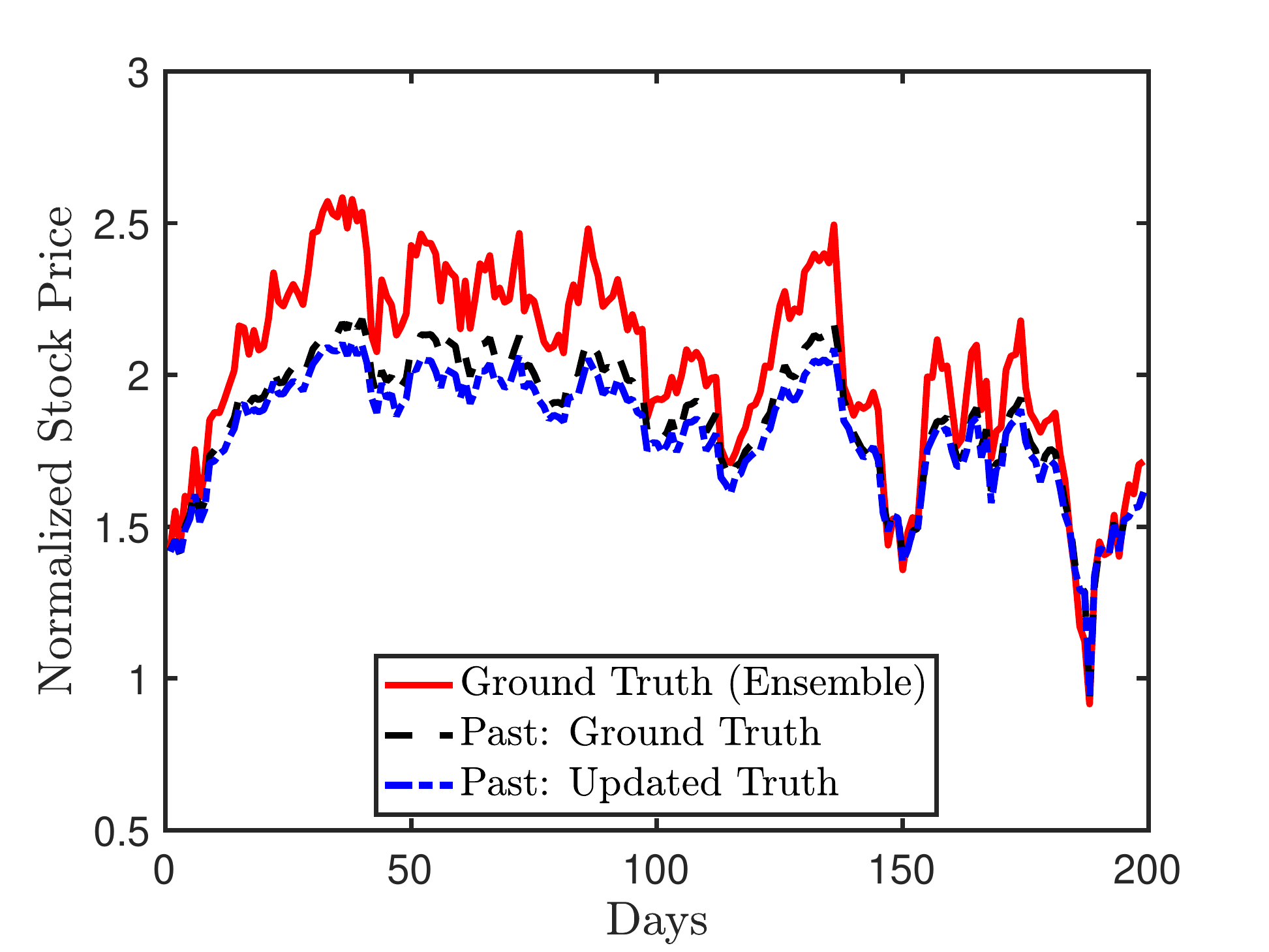}
    \caption{
    (a) Block diagram representing the change in the input and output vector size with the addition of the Google Trends historical data. (b) Predicted values of `XOM' stock using Google Trends and past prices of other stocks in the `energy' industry.   \label{fig:LSTM_GT}}
\end{figure}

The stock price input for this model is of multiple stocks in an industry, identical to the multi-stock model described in Section~\ref{sec:multi_stock}. In Fig.~\ref{fig:LSTM_GT}, we see the difference in the input and output vector sizes in the Block diagram. Additional data, such as more Google Trends keywords and related stock values in the \texttt{axis=-1} direction. Additionally, we also see the result of implementing this model to predict a particular stock's price. A possible reason for the relatively large RMS error is that there is not enough Google Trends data to reduce over-fitting to the training set. Methods such as regularization and adding more keywords related to a single stock can be tried out in the future, to achieve better prediction accuracy.

\section{Conclusions} \label{sec:conclusions}
This paper evaluates candidate state-of-the -art recurrent neural network based architectures to make stock price predictions, namely the stacked LSTM structure and the auto-encoder architecture. On identifying the stacked LSTM architecture as the more effective architecture for stock predictions, we discuss the effect of different dataset preparation techniques on the prediction performance (RMSE). From our dataset preparation tests, we establish that training a model over the historic data available for a collection of stocks allows us to generalize better and make a more accurate forecast. We highlight the utility of the development of such a model to forecast the performance of recently listed stocks which do not have sufficient associated historic data. Next, we demonstrated how a simple decision making bot could be developed to maximize profits by making buy/sell calls on the basis of the stock price prediction obtained from the trained model. The performance of the bot surpasses that of the most successful and aggressive ETFs listed under NYSE, which is to be expected, since maximizing profits requires an accurate mapping of just the qualitative trend of stock prices. A more concerted effort can be made to train models specifically for market crashes or for newly listed stocks in case the generalized model doesn't accurately capture such events. However, we demonstrate a sufficiently accurate prediction model and as a consequence, we were able to make significantly profitable decisions. This shows the effectiveness of using LSTMs for stock price forecasting.

\section*{Author contributions}
N.G. worked on literature study and project planning. S.M. tuned the LSTM architecture and prepared the decision-making bot. A.V. developed all three dataset preparation techniques. S.M., A.V. and N.G. contributed to the writing of the manuscript.

\section*{Conflict of Interest}
The authors declare no competing interests.
\section*{Data Access}
All data is available in the main text. Further information about the computation can be obtained on request to the corresponding author.
\section*{Acknowledgements}
S.M., A.V and N.G. acknowledge the support and guidance from Andrew Ng, Kian Katanforosh, Elane Sui and Vincent Liu.

\end{document}